\documentclass[journal]{IEEEtran}
  \usepackage{graphicx}
  \DeclareGraphicsExtensions{.pdf,.jpeg,.png,.eps}

\usepackage[cmex10]{amsmath}
\usepackage{algorithmic}
\usepackage{array}
\usepackage{mdwmath}
\usepackage{mdwtab}
\usepackage{tabularx}
\usepackage{booktabs}
\usepackage{multirow}
\usepackage{bigstrut}
\usepackage{cite}

\begin{document}
\title{RF-Powered Cognitive Radio Networks: Technical Challenges and Limitations}
\author{Lina Mohjazi, University of Surrey, UK, \\ Mehrdad Dianati, University of Surrey, UK,\\ George K. Karagiannidis, Khalifa University, UAE, and Aristotle University
of Thessaloniki, Greece,\\
Sami Muhaidat\textsuperscript{*}{\thanks{\textsuperscript{*}Corresponding author: Sami Muhaidat, Email: muhaidat@ieee.org}}, Khalifa University, UAE, and University of Surrey, UK,\\
 Mahmoud Al-Qutayri, Khalifa University, UAE.}
       
\markboth{Accepted in IEEE COMMUNICATIONS MAGAZINE}%
{Shell \MakeLowercase{\textit{et al.}}: Bare Demo of IEEEtran.cls for Journals}
\maketitle

\begin{abstract}
The increasing demand for spectral and energy efficient communication networks has spurred a great interest in energy harvesting (EH) cognitive radio networks (CRNs). Such a revolutionary technology represents a paradigm shift in the development of wireless networks, as it can simultaneously enable the efficient use of the available spectrum and the exploitation of radio frequency (RF) energy in order to reduce the reliance on traditional energy sources. This is mainly triggered by the recent advancements in microelectronics that puts forward RF energy harvesting as a plausible technique in the near future. On the other hand, it is suggested that the operation of a network relying on harvested energy needs to be redesigned to allow the network to reliably function in the long term. To this end, the aim of this survey paper is to provide a comprehensive overview of the recent development and the challenges regarding the operation of CRNs powered by RF energy. In addition, the potential open issues that might be considered for the future research are also discussed in this paper.  
 \end{abstract}

\IEEEpeerreviewmaketitle

\section{Introduction}
\label{intro}
Harvesting energy from ambient sources and converting it to electrical energy used to power devices is of increasing importance in designing green communication networks. While this approach enables more environmentally friendly energy supplies, it helps realize the vision for long-lived, self-maintained, and autonomous communication systems. In addition to well-known alternative energy sources, such as solar, wind, geothermal and mechanical, ambient radio-frequency (RF) signals present another promising source that can be exploited in future. A clear advantage of this technique, in comparison with other alternative energy sources, is that ambient RF sources can be consistently available regardless of the time and location in urban areas. Moreover, RF energy harvesting systems can be built cheaply in small dimensions, which could be a significant advantage in the manufacturing of small and low cost communication devices such as sensor nodes. 

\par RF signals can be used by a node to extract information or harvest energy. Scavenging energy from RF signals is broadly known as wireless energy harvesting (EH) or wireless power transfer (WPT), as it refers to the transmission of electrical energy from a power source to one or more electrical loads without any wires. Investigating techniques for RF-powered mobile networks has received significant attention during the past few years in a number of applications such as wireless sensor networks (WSNs), and cooperative communication systems. Most recently, wireless EH has been flagged as a potential source of energy for cognitive radio networks (CRNs) \cite{Seunghyun2013}. The operation of CRNs requires periodical sensing and continuous decision-makings on the availability of spectrum for the secondary users (SUs) in the system. This process along with subsequent signal processing and data transmissions result in high energy consumption by CRN nodes. Thus, it is desirable to find techniques that can help  prolong the lifetime of CRNs. To this end, deploying RF energy harvesting becomes a notable candidate for CRNs, aiming at improving both energy and spectral efficiency of communication networks. In this approach, in addition to the identification of spectrum holes for information transfer, a SU may exploit the ambient RF power to supply an auxiliary source of energy for the CRN nodes. Furthermore, when EH is regarded as the significant source of energy for the operation of CRN nodes, it is crucial that the operation of the system is optimized in order to improve survival of the system, taking into account the characteristics of the considered energy source. This necessitates the need for redesigning of the existing techniques in CRNs in order to simultaneously optimize the EH function and the better utilization of the underlying RF energy source \cite{Lu2014}. 
\par This article aims to review the state-of-the-art of RF-powered CRNs and to survey the enabling techniques that have been proposed in recent years. The remainder of the article is organized as follows. In Section \ref{overview}, the classification of the existing RF energy harvesting techniques are discussed. In Section \ref{arch}, the high level architecture of an RF-powered CRN is presented. This is followed in Section \ref{tech} by surveying the technical aspects that affect the performance of RF-powered CRNs. Furthermore, some of the well-known and promising existing technical solutions in literature are surveyed. Since this research field is still in its early stages, in Section \ref{future}, some of the open technical challenges for possible future investigations are addressed. Finally, concluding remarks are given in Section \ref{conc}.  

\section {Classification of RF Energy Harvesting}
\label {overview}

\par Several methods of WPT have been introduced in the recent literature, including near-field short-range inductive or capacitive coupling, non-radiative mid-range resonance, and far-field long-range RF energy transmission. Nonetheless, the latest class of RF energy transmission in the microwave frequency band is the most recently focused technique. In such frequencies, the wavelength of the RF signal is very small and the WPT system does not require calibration and alignment of the coils and resonators at the transmitter and receiver sides \cite{Shinohara}. This renders the technique as a suitable solution to power a large number of small wireless mobile devices over a wide geographical area. 

Due to the specific communication requirements of the cognitive radio nodes and the nature of RF energy  harvesting, communication techniques and protocols used in the traditional CRNs may not be directly used in RF-powered CRNs \cite{Sungsoo2013}. In particular, it is important to firstly identify the sources of RF energy and their different characteristics in order to understand the technical challenges faced by RF-powered CRNs. The mechanisms by which RF energy is obtained can be mainly classified into two categories: non-intended RF energy harvesting and intended RF energy harvesting. In the following subsections, we provide an overview of these two categories. 

\subsection {Non-intended RF energy harvesting}

Non-intended RF signals are ambient RF sources not originally intended for energy transfer. This includes signals radiated due to wireless telecommunication services, such as cellular systems, mobile devices, and Wireless Local Area Networks (WLANs), or from public broadcasting systems, such as TV and radio. These ambient signals, if not received by their intended receivers, are dissipated as heat, resulting in a waste of energy. Instead, they could be used as a sustainable and low-cost source to harvest energy from \cite{Valenta}. A device that harvests energy from ambient RF sources can have separate antennas or antenna array for RF transceiver and RF energy harvester. Harvesting energy by this means is subject to long-term and short-term fluctuations due to radio tower service schedules, nodes mobility and activity patterns, and fading. Therefore, cognitive radio terminals should employ new schemes that consider the tradeoff among network throughput, energy efficiency, and RF energy supply, given the dynamic availability of the RF energy.      

\subsection {Intended RF energy harvesting}

This method can be divided into two types. In the first, the receiver obtains wireless power transferred from a dedicated source that only delivers power without transmitting information to it such as directive power beamforming\footnote{Powercast transmitter is one example that is already commercialized. Interested readers may learn more at http://www.powercastco.com/}. The second method uses the same emitted RF signal to transport energy and information simultaneously, known as simultaneous wireless information and power transfer (SWIPT) \cite{Varshney2008}. 
\par A number of receiver designs have been proposed for SWIPT. The two most adopted designs in literature are the integrated and the co-located receiver design. The co-located receiver design can be based on either time switching or power splitting \cite{ZhangMay2013}. A power splitting block divides the received signal into two portions, one for EH and the other for information decoding, while time switching allocates dedicated time slots to EH and the rest for data processing. By employing this approach, controllable and efficient on-demand wireless information and energy can be simultaneously provided. This permits a low-cost alternative for sustainable wireless systems without further hardware modification on the transmitter side.

\section{Overview of RF-powered CRNs}
\label{arch}

\par
\begin{figure*}[!t]
\centering
\includegraphics[width=6in]{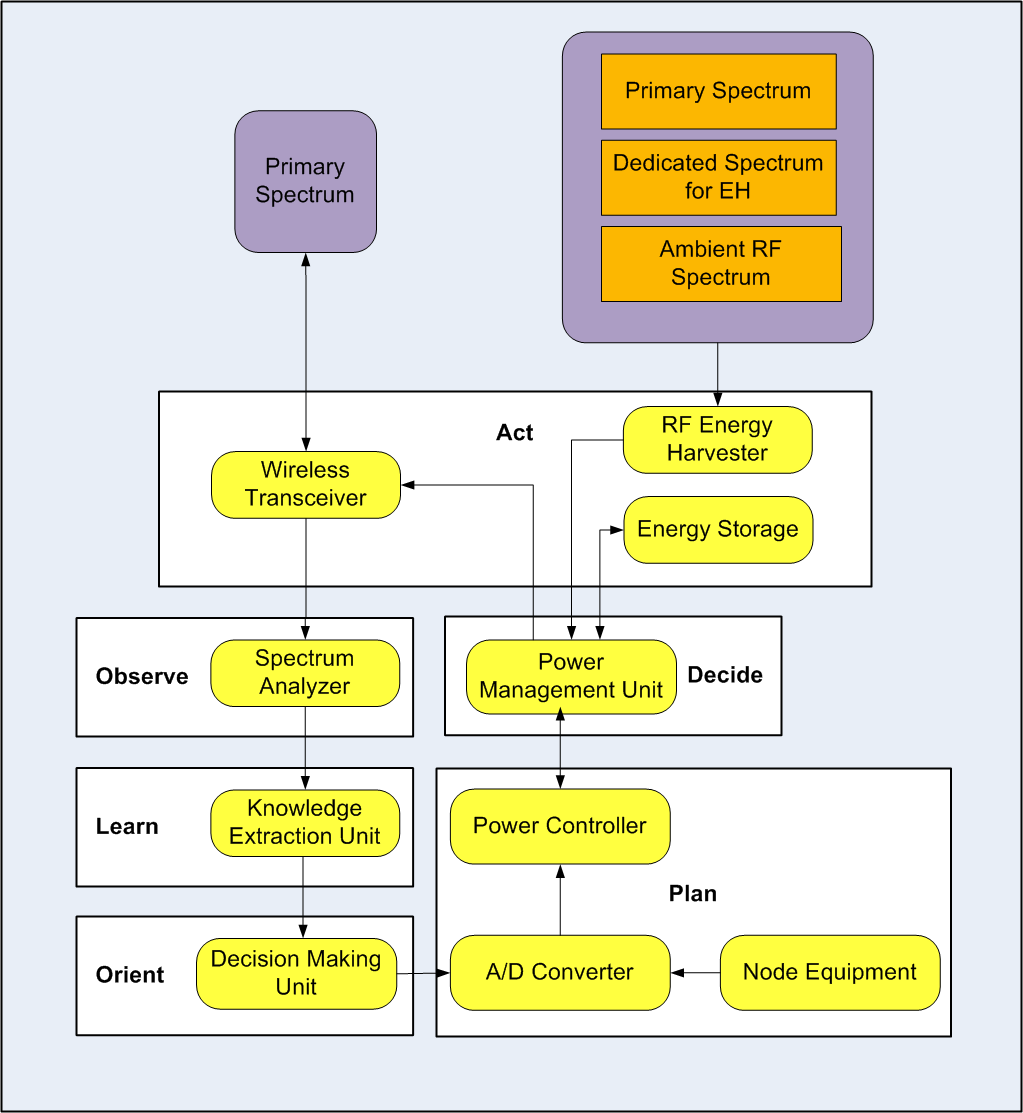}
\caption{RF-Powered CRN Node Operation Cycle Block Diagram \cite{Lu2014}}
\label{CRdevice}
\end{figure*}

 There has been recent interest in exploitation of RF based EH for CRNs. As it is the main focus of this paper, in the following, we elaborate on this application with further details. A general block diagram of the functions performed by a cognitive radio node with RF energy harvesting capability is illustrated in Fig. \ref{CRdevice} \cite{Lu2014}. The role of each component is described related to the major functions of a cognitive cycle, i.e., observing, learning, orienting, planning, deciding and acting, as follows: 
\begin{itemize}
	\item Wireless transceiver: a software-defined radio for data transmission and reception.
	\item Energy storage: this could be a battery or capacitor to store the harvested energy.
	\item Power management unit:  decides whether the harvested energy should be stored in energy storage or forwarded to other components.
	\item RF energy harvester: replenishes RF signals and converts them to electricity.
	\item Spectrum analyzer: provides instantaneous analysis of the activity of spectrum usage.
	\item Knowledge extraction unit: maintains a record about the spectrum access environment.
	\item Decision making unit: decides on spectrum access.
	\item Node equipment: implements device applications.
	\item A/D converter: digitizes the analog signal produced by the node equipment.
	\item Power controller: processes the output of the A/D converter for network applications.
\end{itemize}

\begin{figure*}[!t]
\centering
\includegraphics[width=6in]{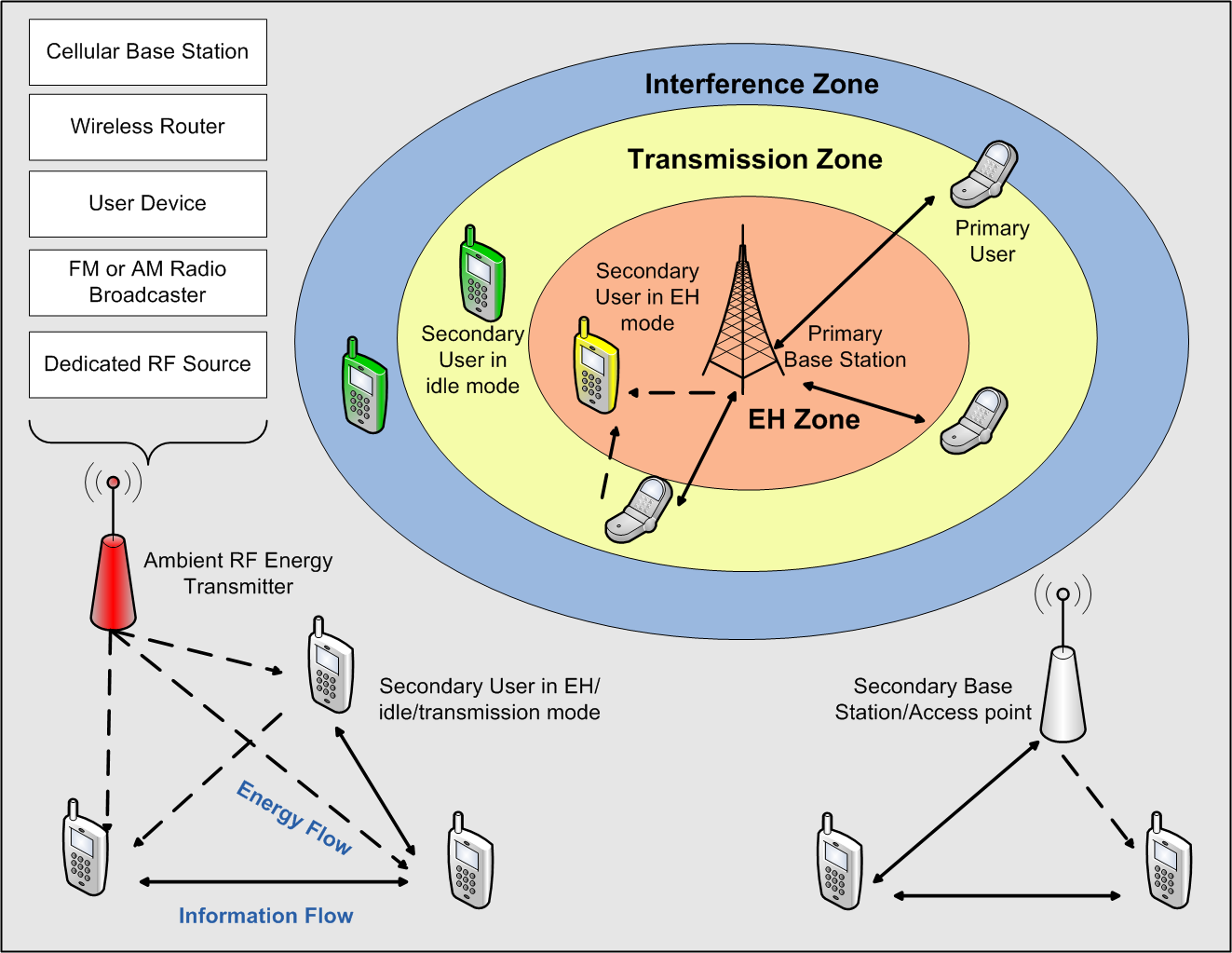}
\caption{A general architecture of an RF-powered CRN}
\label{network}
\end{figure*}

A general architecture of CRN powered by either ambient RF signals, energy transmitted from an intended RF source or via SWIPT is shown in Fig. \ref{network}. When SUs harvest RF energy from the primary network, the primary base station can be associated with three zones \cite{Seunghyun2013} that defines the SUs activity. Secondary users that are not fully charged and are located in the EH zone can harvest energy from the RF signals received from the primary base station or nearby PUs. SUs which are located inside the interference zone can not transmit unless the spectrum is unoccupied by the PUs. Furthermore, it can be seen from Fig. \ref{network} that the secondary network can also harvest ambient RF energy. RF-powered CRNs can adopt either an infrastructure-based or an infrastructure-less communication architecture.

\section{Technical Challenges of RF-powered CRNs}
\label{tech}

As discussed in the previous sections, CRN nodes may be powered by two different categories of RF energy sources. In this section, we provide an overview on the technical challenges arise in both scenarios. 
\par In the scenario where a cognitive radio node harvest energy from non-intended RF energy, the energy available randomly varies over the time, i.e., a random process, known as the energy prof\mbox{}ile, which can be described by certain mathematical models. This inherent randomness of the energy source is a major factor that affects the performance of an EH node. On the other hand, an SU can also receive RF energy either from ambient transmissions of the primary network or from a particular PU with activity known to the SU. In this case, the cognitive operation of the SU is powered solely by the RF energy from the PU. Therefore, both the occupied and the idle spectrum are essential for the operation of a SU. In both the aforementioned cases, the performance of a CRN is restricted by the \textit{collision constraint} which requires that the probability of colliding with the primary transmission is always kept below a predef\mbox{}ined threshold. When a SU operates in a time-slotted manner, its frame structure is divided into several time slots to perform different cognitive radio tasks. The performance of each of them is directly affected by the available energy at the time when it is to be executed. The total consumed energy should be equal to or less than the total harvested energy, this is called the\textit{EH constraint} \cite{Seunghyun2013}. Putting those two constraints together imply fundamental limitations on the throughput of an EH CRN.
\par Several studies focused on exploring the impact of EH on CRNs. A seminal work in this area is \cite{Seunghyun2013} that proposes a novel framework, enabling SUs to opportunistically harvest ambient RF energy as well as reuse the spectrum of PUs. Also, the transmission probability of SUs and the resulting system throughput of the CRN were derived when a stochastic-geometry model is considered. The results presented in \cite{Seunghyun2013} revealed key insights about the optimal network design. Moreover, the authors in \cite{Daesik} derived the upper bound on the achievable throughput as a function of the energy arrival rate, the temporal correlation of the primary traff\mbox{}ic, and the detection threshold for a spectrum sensor. 
\par We aim in this section to discuss techniques that should be revisited in order to optimize system conf\mbox{}igurations to accommodate for the newly introduced requirements of RF-powered CRNS. In addition, we review the relevant solutions proposed in literature.

\subsection {Mode Selection}

A SU harvesting ambient RF energy usually operates either in an active or a sleep mode. In the former, it performs spectrum sensing and then data transmission, if the detector decides that the primary user is absent. In the latter, the SU remains silent and only harvests energy. On the other hard, when a SU needs to exploit the existence of the PU to harvest RF energy, it selects either the spectrum access mode (including sensing the idle spectrum then transmission, or sensing the occupied spectrum then harvesting) or the harvesting mode that only incorporates the process of EH. There is a trade-off for each node between utilization of the spectrum and exploitation of RF energy. The more a node spends time to sense spectrum holes and use the opportunities for transmission the higher is the energy consumption rate and the fewer the opportunities for EH. Therefore, in order to simultaneously enhance the network performance and the energy utilization, an optimal mode selection policy may be investigated. Motivated by this trade-off, the work in \cite{Park2012} considers a cognitive radio sensor network where SUs performs either RF energy harvesting or opportunistic spectrum access at a time. Under this assumption, the authors developed an optimal mode selection policy in the framework of a partially observable Markov decision process (POMDP). Built on the concept of hybrid underlay-overlay spectrum access, the work carried out in \cite{Usman2014} proposed a mode selection strategy where the SU can be in one of three states, i.e., transmission mode (either underlay or overlay), sleep mode, EH mode. The objective is to f\mbox{}ind a balance between the system throughput and the harvested energy for future use. 
\par Since the transmitted power attenuates according to the reciprocal of the distance, to ensure a certain EH eff\mbox{}iciency, the decision to select the harvesting mode has to consider both the availability of the PU and its distance from the SU, as studied in \cite{Seunghyun2013}.

\subsection {Sensing Duration} 
	
The main question here is to determine how the duration of spectrum access is constrained by the sensing process, which is crucial to system performance. Longer sensing duration results in higher probability of true detections of the spectrum and thus lower interference caused to PUs. However, it simultaneously decreases the chances of the SU in accessing the spectrum. The total energy consumption behavior varies from one frame to the other according to the variation in the sensing duration. This behavior not only depends on the sensing duration, it is also affected by the sensing-to-transmission power ratio. Both, the opportunities of accessing the idle spectrum and the energy consumed by sensing increase as the sensing duration increases. This also elevates the energy consumed by more frequent data transmissions. Nonetheless, if the sensing duration is too long, the time left for transmission becomes short and accordingly, the total amount of energy consumption (sensing plus transmission energies) is reduced, due to the decreased opportunity of data transmission. The aforementioned conflicting factors collectively imply coming up with an optimal sensing duration, that would take into account the available energy and the effect on the performance of both CR and primary networks. In \cite {Sixing} for example, the authors derived a mechanism that jointly optimizes the harvesting, sensing, and transmitting durations and the number of sensed channels based on mixed-integer non-linear programming with maximizing the achievable throughput serving as the objective function. Recently, the study of \cite{Wonsuk2014} suggested a new policy for determining both the sensing duration and the detection threshold that maximizes the average throughput. The proposed technique aims to f\mbox{}ind an optimal pair of sensing duration and detection threshold that can increase the spectrum access opportunities within the permissible range of collision probability for a given average harvested energy.

\subsection {Detection Threshold}
	
The performance of detecting the existence of primary signals is linked to the chosen value of the detection threshold. The choice of this value becomes even more crucial when the SU is an EH node \cite{Sungsoo2013}. In general, a high detection threshold increases the probability of detecting the spectrum as idle and leads to more frequent spectrum access. This does not only increase the probability of colliding with the PU transmissions, but also cause a large waste of energy resulting from more transmissions. On the contrary, a low detection threshold alleviates unnecessary energy waste and the probability of accessing the occupied spectrum, but may in turn refrains the SU from transmitting data, even when the spectrum is idle. In \cite{Sungsoo2013}, the authors propose a technique by which an optimal detection threshold is derived, using the probability of accessing the idle spectrum and the probability of accessing the occupied spectrum to maximize the expected total throughput while satisfying both the EH and the collision constraints. They have also demonstrated that, depending on the selected threshold, the system can be characterized as a \textit{spectrum-limited regime} and an \textit{energy-limited regime}. In the first, the harvested energy enables continuous spectrum access, while in the second, the amount of harvested energy restricts the number of spectrum access attempts. This work was followed by the one presented in \cite{SPark2013} where they have extended the problem in \cite {Sungsoo2013} to a joint optimization problem of a spectrum sensing policy and a detection threshold subject to the EH and collision constraints. In the framework of a POMDP, this strategy is able to achieve eff\mbox{}icient usage of the harvested energy by exploiting the temporal correlation of the primary traff\mbox{}ic. In addition to deriving the upper bound on the achievable throughput in \cite{Daesik}, the authors have also explored a new technique that is able to f\mbox{}ind the optimal detection threshold that maximizes the derived upper bound.

\par If a SU employs SWIPT in order to simultaneously use the received RF signal to store energy and detect the presence of the PU, it is challenging to choose the optimal detection threshold. For example, in the power splitting approach, where the received signal at the SU is split into two portions, one for EH and the other for energy detection, the value of the detection threshold used in a non-EH SU receiver will not be viable. The reason is that, the minimum acceptable signal energy at the input of the energy detector is divided according to the power splitting ratio. Hence, the detection threshold should correspond to the value of the received power after being split. This raises the question about the choice of the energy threshold in the occasion where the power splitting ratio is varying.

\subsection {Energy Management}

A careful allocation of power over sensing and data transmission slots is of high importance, due to its effect on the system throughput, capacity, and outage probability. In a CRN powered by ambient RF energy, the energy available at the beginning of a time slot is divided between the spectrum sensing and data transmission phases. Therefore, the harvested energy has to be eff\mbox{}iciently expended over a specif\mbox{}ic number of time slots, in order to enhance the system performance. The mechanism proposed in \cite{sultan2012}, for instance, enables an EH cognitive radio node to optimize its sensing and transmit energies while accounting for the detection reliability-throughput tradeoff. Another method to achieve energy management is via the knowledge of the previous or current statistics of the energy arrival rate, the statistical description of PUs's activity, or the channel state information (CSI). For example, in \cite {Gao}, the proposed scheme allocates more energy for transmission when the channel state is good in a particular time slot. In contrast, less or no energy is allocated to a transmission slot, in which the probability that the PU occupies the spectrum is anticipated to be relatively high. 

\par The problem of energy management in a CRN applying SWIPT differs substantially from the one that harvests ambient RF energy. The reason is that, in some scenarios in  SWIPT, the receiver has no battery to store energy, and as a result, the processes to be executed in a certain time slot directly draw energy from the one available by the received RF signal. In this situation, it is challenging to optimize the parameters of the SU receiver, such that energy is distributed spontaneously and eff\mbox{}iciently between the different tasks of the cognitive cycle.

\subsection{Channel Selection}

Traditional channel selection schemes, which mainly aim at identifying the idle channels with high quality, may not be effective anymore for RF-powered CRNs. In particular, if the energy level available at the SU is low, it might select the channel which tends to be occupied by a PU and has a strong RF signal to harvest. On the other hand, if the SU has a high energy level and there is a need for data packets transmission, it should identify the channel, which is likely to be idle with a favorable channel quality. The research work reported in \cite {pradha} studied a channel selection criterion that maximizes the average spectral eff\mbox{}iciency of a SU. The proposed method jointly exploits the knowledge of the PU occupancy and channel conditions, and the dependency of the decision of the SU to sense and access the PU spectrum on the probabilistic availability of energy at the SU. Similarly, in \cite{Lu2014}, the authors developed a channel selection policy used by the SU that maps the SU's state (i.e. number of packets in the data queue and the energy level of the energy storage) to the channel to be selected. This is done prior to sensing the channel and is based on statistical information such as probabilities of channel to be idle and busy, the probability of successful packet transmission if the channel is idle, and the probability of successful EH if the channel is busy.

\par Table \ref{summary} shows a summary of existing conf\mbox{}iguration policies for RF-powered CRNs.

\begin{table*}[!t]
  \centering
  \caption{Summary of Proposed Techniques for RF-Powered CRNs}
	
    \begin{tabular}{|>{\centering\arraybackslash}m{0.9in}|>{\centering\arraybackslash}m{0.5in}|>{\centering\arraybackslash}m{0.8in}|>{\centering\arraybackslash}m{1.6in}|>{\arraybackslash}m{1.2in}|>{\centering\arraybackslash}m{1.2in}|}
    \toprule
 \textbf{Configuration Element} & \textbf{Literature} & \textbf{EH Model} & \textbf{Constraints} & \centering{\textbf{Objective}}& \textbf{Framework}\\ 
    \midrule 
    \multirow{2}[4]{*}{Mode Selection} & \cite{Park2012} & \parbox[l]{2cm}{Opportunistic EH of RF signals from primary network} & \parbox[l]{4cm}{1) Residual energy at the SU\\2) Spectrum occupancy state partially observable to the sensor node} & {Maximize expected total throughput delivered by a SU sensor node over a time slot} & \parbox[c]{3cm}{POMDP} \bigstrut \\ \cline{2-6}

& \cite{Usman2014}  & \parbox[l]{2cm}{Stochastic EH of RF signals from primary network and ambient RF sources} & \parbox[l]{4cm}{1) Residual energy at the SU \\2) Required transmission energy\\3) Spectrum occupancy state partially observable} & {Enhance throughput of the SU and obtain QoS of primary network by selecting overlay or underlay transmission mode} & \parbox[c]{3cm}{POMDP}\bigstrut \\ \hline

\multirow{2}[4]{*}{Sensing Duration} & \cite{Sixing} & \parbox[l]{2cm}{EH from ambient RF sources} & \parbox[l]{4cm}{1) EH rate of the SU\\2) Collision constraint to the primary network\\3) Channel sensing energy cost} & {Optimize saving-sensing-transmitting structure that maximizes the achievable throughput of the SU} & \parbox[c]{3cm}{Mixed-integer non-linear programming} \bigstrut \\ \cline{2-6}

& \cite{Wonsuk2014} & \parbox[l]{2cm}{EH from ambient RF and other energy sources} & \parbox[l]{4cm}{1) Channel sensing and data transmission energy cost with respect to the residual energy at the SU\\2) Collision constraint to the primary network} & {Maximize expected average throughput of the secondary network} & \parbox[l]{3cm}{Several optimization problems are formulated to give an insight on the joint conf\mbox{}iguration of sensing duration and threshold}\bigstrut \\ \hline

  \multirow{3}[6]{*}{Detection Threshold} & \cite{Sungsoo2013} & \parbox[l]{2cm}{EH from ambient RF and other energy sources} & \parbox[l]{4cm}{1) Energy arrival rate\\2) Channel sensing and data transmission energy cost with respect to the residual energy at the SU\\3) Collision constraint to the primary network} & {Maximize expected total throughput of the secondary network} & \parbox[l]{3cm}{Deriving the probability of accessing the idle spectrum and the probability of accessing the occupied spectrum and their bounds}\bigstrut \\ \cline{2-6}

& \cite{SPark2013} & \parbox[l]{2cm}{EH from ambient RF and other energy sources} & \parbox[l]{4cm}{1) Spectrum occupancy state partially observable\\
2) Energy arrival rate\\3) Temporal correlation of the primary traffic\\4) Collision constraint to the primary network} & {Maximize the upper bound of the probability of accessing the idle spectrum} & \parbox[l]{3cm}{Unconstrained POMDP}\bigstrut \\ \cline{2-6}

  & \cite{Daesik} & \parbox[l]{2cm}{EH from ambient RF and other energy sources} & \parbox[l]{4cm}{1) Energy arrival rate\\ 2) Channel sensing and data transmission energy cost with respect to the residual energy at the SU\\3) Temporal correlation of the primary traff\mbox{}ic\\4) Collision constraint to the primary network} & {Maximize the upper bound of the achievable throughput} & \parbox[l]{3cm}{Several optimization problems are formulated to give an insight on the joint configuration of spectrum access policy and detection threshold}\bigstrut \\ \hline
 
\multirow{2}[4]{*}{Energy Management} & \cite{sultan2012} & \parbox[l]{2cm}{EH from ambient RF and other energy sources} & \parbox[l]{4cm}{1) Energy arrival rate\\2) Residual energy at the SU} & {Maximize expected total throughput of the secondary network} & \parbox[l]{3cm}{Markovian decision process}\bigstrut \\ \cline{2-6}

& \cite {Gao} & \parbox[l]{2cm}{EH from ambient RF and other energy sources} & \parbox[l]{4cm}{1) Observed information (harvested energy, fading CSI, spectrum occupancy state ) in the past and present only} & {Maximize expected total throughput of the secondary network} & \parbox[l]{3cm}{Sliding window approach}\bigstrut \\ \hline

    \multirow{2}[4]{*}{Channel Selection} & \cite {pradha}  & \parbox[l]{2cm}{EH from ambient RF and other energy sources} & \parbox[l]{4cm}{1) Probabilistic availability of energy at SU\\2) Channel conditions\\3) Primary network belief state} & {Maximize expected total throughput of the secondary network} & \parbox[l]{3cm}{POMDP}\bigstrut \\ \cline{2-6}
		
		          & \cite{Lu2014} & \parbox[l]{2cm}{EH from RF signals of  primary network} & \parbox[l]{4cm}{1) Number of packets in the data queue\\2) Residual energy at the SU} & {Maximize the long-term average throughput of the SU} & \parbox[l]{3cm}{Markovian decision process} \\

    \bottomrule
    \end{tabular}
  \label{summary}
\end{table*}

\section {Future Research for RF-Powered CRNs}  
\label{future}   
CRNs may be deployed in different scenarios such as multiple-input multiple-output (MIMO), cooperative, and relaying CRNs. Existing mechanisms for conventional CRNs need to be extended, modif\mbox{}ied, or even replaced to suit the newly emerged RF-based EH technology. We focus next on discussing some issues that can be explored in future. 

\subsection {Sensing Imperfections}
Protecting the primary network from unbearable interference is the key to a successful operation of a CRN. Therefore, a high probability of correct decisions generated by the energy detector, is vital. In practice, however, those decisions are prone to errors leading the performance of the primary network and the CRN to dramatically deteriorate. This becomes of a higher concern in the presence of EH in those networks. In particular, if the channel was sensed as idle, while it is actually busy, and if SU decides to transmit, this results in unnecessary dissipation of energy, causing interference to the PU, and missing a chance to harvest energy if needed. On the other hand, if the channel was sensed as busy, while it is in fact idle, the SU might preserve energy but it abolishes an opportunity to provide a better rate to its intended receiver. This necessitate research studies to explore the limitations caused by imperfect sensing on the performance of RF-based EH CRNs.

\subsection {CRNs with multiple antennas}
Multiple antennas in CRNs can be utilized to provide the secondary transmitter with more degrees of freedom in space in addition to time and frequency. Multi-antenna CRNs gained attraction specially in the underlay spectrum sharing scheme, where SU and PU transmissions can be concurrent. In line with this, it is known that higher wireless energy transfer eff\mbox{}iciencies can be achieved when multiple antennas are employed. Furthermore, in a multi-antenna RF-powered CRN, beamforming techniques can be exploited by the SU transmitter to steer RF signals towards SU receivers having different information and/or EH requirements. The problem of maximizing the SU rate subject to both the PU rate and the secondary transmitter power constraints is critical. Therefore, beamforming techniques should be redesigned to consider those conflicting objectives. The work presented in \cite{Zheng} is a major development in this f\mbox{}ield, where a multi-antenna EH secondary network makes use of both the spectrum and the energy of the primary network, in return to assist the primary transmissions. The main focus of this research is to design a beamforming technique that characterizes the achievable primary-secondary rate region based on power splitting and time-switching for SWIPT. 
\par Beamforming performance optimization is tightly dependent on the acquisition of CSI. As a result, new mechanisms have to be proposed to account for the tradeoff between data transmission, EH, and channel state estimation duration.

\subsection{Cooperative CRNs}
The concept of cooperative spectrum sensing has been proven to combat sensing errors and channel fading, and to overcome the hidden terminal problem due to shadowing. Nevertheless, conventional cooperative schemes do not take into consideration the DC power levels produced by the RF energy conversion process, which resemble the only source of energy available at the CR terminal. To be more specif\mbox{}ic, a SU might refrain from participating in the process of spectrum sensing because it does not receive suff\mbox{}icient RF energy due its proximity from the PU. However, the more SUs that participate in sensing the better spectrum discovery outcome is guaranteed and the more energy will be consumed. As a consequence, centralized cooperative spectrum scheduling, in which a cognitive base station or a fusion center decides which SUs should participate in the sensing process and which channels to sense, should take into account the amounts of the harvested energy at the SUs. In addition, the distances between a PU transmitter and different SUs are often different. Also, the signal propagation environment differs from a PU transmitter to different SUs, making both the signal-to-noise ratio (SNR) and the harvested energy from the same primary signal dissimilar at different SU receivers. Therefore, new cooperative mechanisms that f\mbox{}it into this environment is thus essential.

\subsection{CRNs with relays}
In a cognitive relaying network, a single or multiple relays assist the SU source to sense and/or transmit data to the SU destination. All the CRN nodes or only the relay/s might be RF-based EH. In the second scenario, relays harvest energy from either the SU source, or the PU, or both. Under this setting, the quality of relaying the data to the SU destination is directly affected by the power received at the relay/s from the SU source or the PU signals. This problem seems to be even more complex if the relay/s and the SU source deploy SWIPT. In such a case, both the SU source and the relay/s have to precisely select their receiver parameters (power splitting or time switching ratios) in order to optimize the overall system performance, while satisfying their energy needs. As a consequence, more research focus has to be directed towards exploring new relaying protocols and relay selection schemes.                 

\section{Conclusions}
\label{conc}
The recent interest in simultaneously achieving spectrum and energy eff\mbox{}iciency has led to the concept of RF-powered CRNs. Integrating the capability of EH into the functionality of cognitive radio devices infer nontrivial challenges on their designs. This article presented an overview of the architecture of CRNs that operate based on RF energy harvesting. Mainly, two methods by which CRNs can harvest RF energy were discussed: intended and non-intended RF energy harvesting. Several factors that do not exist in non-RF-powered CRNs impose fundamental limitations on their performance. As a result, the paper listed key conf\mbox{}iguration parameters that need to be redesigned to achieve a desirable balance between the energy availability constraint and the system performance. Furthermore, the article surveyed promising techniques that can enable successful spectrum sensing, spectrum access, and spectrum management in RF-powered CRNs. Finally, some open technical challenges that may be studied in future were addressed.

\bibliographystyle{IEEEtran}
\bstctlcite{BSTcontrol}
\bibliography{Reflist}

\begin{IEEEbiographynophoto}
{Lina Mohjazi}(l.mohjazi@surrey.ac.uk) received a B.Eng degree in electrical and electronic/communication engineering from the UAE University, UAE, in 2008, and her M.Sc. by research degree in communications engineering at Khalifa University, UAE, in 2012. Since October 2013, she has been a Ph.D. student at the University of Surrey in United Kingdom. Her main research interests include cognitive radio networks, energy harvesting communication systems, and physical layer optimization.
\end{IEEEbiographynophoto}
\begin{IEEEbiographynophoto}
{Mehrdad Dianati} (m.dianati@surrey.ac.uk) a Reader (Associate Professor) in Communication and Networking Systems at the Institute of Communication Systems (ICS) of the University of Surrey in United Kingdom. His research area mainly includes wireless access networks and connected/autonomous vehicles. Mehrdad also has 9 years of industrial experience as software/hardware developer and Director of R\&D. He is currently an associate editor for IEEE Transactions on Vehicular Technology, IET Communications and Wiley's Journal of Wireless Communications and Mobile.
\end{IEEEbiographynophoto}
\begin{IEEEbiographynophoto}
{George K. Karagiannidis} (geokarag@ieee.org) George K. Karagiannidis received a PhD degree in ECE from the University of Patras, in 1999. In 2004, he joined the faculty of Aristotle University of Thessaloniki, Greece, where he is Professor in the ECE Dept. and Director of Digital Telecommunications Systems and Networks Laboratory. In 2014, he joined Khalifa University, UAE, where is currently Professor in the ECE Dept. and Coordinator of the ICT Cluster. Since January 2012 he is the Editor-in Chief of IEEE Communications Letters.
\end{IEEEbiographynophoto}
\begin{IEEEbiographynophoto}
{Sami Muhaidat} (muhaidat@ieee.org) received the Ph.D. degree in Electrical and Computer Engineering from the University of Waterloo, Canada.  He is currently an Associate Professor at Khalifa University and a Visiting Professor in the department of Electrical and Computer Engineering, University of Western Ontario, Canada. Sami currently serves as an Editor for IEEE Communications Letters and an Associate Editor for IEEE Transactions on Vehicular Technology.
\end{IEEEbiographynophoto}
\begin{IEEEbiographynophoto}
{Mahmoud Al-Qutayri} (mqutayri@kustar.ac.ae) is a Professor in the Department of Electrical and Computer Engineering and the Associate Dean for Graduate Studies at Khalifa University, UAE.  He received the B.Eng., MSc and PhD degrees from Concordia University, Canada, University of Manchester, U.K., and the University of Bath, U.K., all in Electrical and Electronic Engineering in 1984, 1987, and 1992, respectively. He has published numerous technical papers in peer reviewed international journals and conferences, and coauthored a book. His fields of research include embedded systems, wireless sensor networks, cognitive radio, and mixed-signal circuits. 
\end{IEEEbiographynophoto}

\end{document}